\documentclass[runningheads]{llncs}
\usepackage[T1]{fontenc}
\usepackage{graphicx}
\usepackage{tabularx}
\usepackage{adjustbox}
\usepackage{float}
\usepackage{makecell}
\usepackage{amsfonts}
\usepackage{enumitem}
\usepackage{capt-of}
\usepackage{times}
\usepackage{latexsym}
\usepackage{amsmath}
\usepackage{amssymb}
\begin{document}
\title{Personalized Product Search Ranking: A Multi-Task Learning Approach with Tabular and Non-Tabular Data}
\titlerunning{Personalized Product Search Ranking}
%

\author{
Lalitesh Morishetti\inst{1}\thanks{These authors contributed equally to this work.} \and
Abhay Kumar\inst{1}\textsuperscript{$\star$} \and
Jonathan Scott\inst{1} \and
Kaushiki Nag\inst{1} \and
Gunjan Sharma\inst{2} \and
Shanu Vashishtha\inst{1} \and
Rahul Sridhar\inst{2} \and
Rohit Chatter\inst{2} \and
Kannan Achan\inst{1}
}

\authorrunning{ }
\institute{Personalization and Recommendation, Walmart Global Tech, Sunnyvale, USA \and
Personalization and Recommendation, Walmart Global Tech, Bengaluru, India\\
\email{\{lalitesh.morishetti,abhay.kumar,jonathan.scott,kaushiki.nag, gunjan.sharma4, Shanu.Vashishtha, rahul.sridhar, rohit.chatter, kannan.achan\}@walmart.com}}
\maketitle              
\begin{abstract}
In this paper, we present a novel model architecture for optimizing personalized product search ranking using a multi-task learning (MTL) framework. Our approach uniquely integrates tabular and non-tabular data, leveraging a pre-trained TinyBERT\cite{jiao2020tinybertdistillingbertnatural} model for semantic embeddings and a novel sampling technique to capture diverse customer behaviors. We evaluate our model against several baselines, including XGBoost\cite{chen2016xgboost}, TabNet\cite{arik2021tabnet}, FT-Transformer\cite{gorishniy2021revisiting}, DCN-V2\cite{wang2021dcn}, and MMoE\cite{ma2018modeling}, focusing on their ability to handle mixed data types and optimize personalized ranking. Additionally, we propose a scalable relevance labeling mechanism based on click-through rates, click positions, and semantic similarity, offering an alternative to traditional human-annotated labels. Experimental results show that combining non-tabular data with advanced embedding techniques in multi-task learning paradigm significantly enhances model performance. Ablation studies further underscore the benefits of incorporating relevance labels, fine-tuning TinyBERT layers, and TinyBERT query-product embedding interactions. These results demonstrate the effectiveness of our approach in achieving improved personalized product search ranking.

\keywords{personalized product search, multi-task learning, recommendation systems, product ranking, semantic embeddings, e-commerce}
\end{abstract}
\section{Introduction}
In the dynamic landscape of e-commerce, personalized product search ranking is essential for enhancing user experience and driving conversions. The goal is to present users with search results that are not only relevant to their queries but also aligned with their individual preferences and behaviors. Traditional search ranking models, such as those based on tree ensembles or deep learning, have primarily focused on optimizing single tasks \cite{burges2010from,INR-016}. These models are increasingly insufficient for handling the complexity of modern search engines, where multiple objectives such as query relevance, abandonment rate \cite{10.1145/2983323.2983867}, and user engagement \cite{10.1145/2556195.2556220} must be balanced. Additionally, most popular ranking models are tree-based models \cite{pmlr-v14-chapelle11a,ke2017lightgbm} which are limited to tabular data and often struggle to capture the complex interactions between user intent, product attributes, and contextual factors, particularly when these factors span diverse data types.

To address these challenges, we propose a novel multi-task learning (MTL) framework that integrates both tabular and non-tabular data to optimize personalized product search ranking. While tabular data, comprising features like user demographics, click history, and product attributes, is essential for personalization, non-tabular data, such as textual descriptions and user reviews, adds critical context that can improve the relevance of search results. Our approach leverages a combination of deep neural networks and pre-trained language models to enhance the model’s ability to learn from diverse data sources. By incorporating semantic embeddings extracted from a fine-tuned TinyBERT model, along with traditional tabular features, our framework captures richer representations of query-product relevance. Furthermore, we introduce a relevance labeling mechanism based on a voting scheme that combines click-through rate, click position, and semantic similarity, providing a robust alternative to costly human-annotated relevance labels.
This paper makes the following key contributions:
\begin{itemize}
    \item \textbf{Novel MTL Framework for Product Search Ranking}: We propose a multi-task learning architecture that effectively integrates both tabular and non-tabular data, leveraging a pre-trained TinyBERT model for semantic embedding extraction to improve query-item relevance.
    \item \textbf{Relevance Labeling Mechanism}: We introduce a scalable relevance labeling technique that combines multiple signals—click-through rate, click position, and semantic similarity scores—to generate relevance labels. This method provides a cost-effective alternative to human-annotated labels while maintaining robustness against noise in click logs.
    \item \textbf{Comprehensive Comparative Study}: We conduct a detailed comparative analysis of several baseline models, including XGBoost\cite{chen2016xgboost}, TabNet\cite{arik2021tabnet}, Feature Tokenizer + Transformer (FT-T)\cite{gorishniy2021revisiting}, DCN-V2\cite{wang2021dcn}, and Multi-gate Mixture-of-Experts (MMoE)\cite{ma2018modeling}, to evaluate their effectiveness in handling different data types and optimizing personalized ranking tasks.
\end{itemize}
We perform extensive experimental validation, demonstrating the superiority of our proposed framework in capturing complex interactions and delivering improved ranking performance across multiple tasks. The results highlight the significance of combining tabular and non-tabular data for enhancing the personalization of product search results.

\section{Related Works}

The literature on search ranking systems is extensive and covers various approaches. Our work is primarily related to the ranking in e-commerce product setup. In this section, we review works closely related to our proposed system such as multi-task learning, learning to rank and deep neural networks designed for tabular data. Common recommendation tasks include click through rate prediction. Popular approaches to search product ranking include neighborhood based matrix factorization methods \cite{xiang2010context,ning2015comprehensive,frolov2017tensor}, more recent deep learning based methods\cite{ijcai2017p239,10.1145/3219819.3219823,bi2019leverage}  and deep cross networks \cite{wang2017deep,naumov2019deep,wang2021dcn}. These methods are single task methods but in practice a user does multiple actions (especially within an e-commerce context) such as click, add-to-cart or checkout (complete a transaction). Due to these multi-action user sessions, developing models based on one task is not sufficient. 

In recent years, multi-task recommendation deep learning based modeling has become quite popular. Due to the flexibility and capability of deep neural networks in learning high-order feature interactions and complex user-item interactions, these deep learning based multi-task based systems have demonstrated superior performance compared to traditional multi-task recommendation models\cite{Wu_2022}. Tasks can be categorized according to \cite{wang2023multi}: \textbf{parallel:} there is no sequential dependence between tasks and the objective function is usually defined as the weighted sum of losses for each task, \textbf{cascaded:} refers to there being sequential dependence between tasks, and lastly \textbf{auxiliary:} where an additional task helps improve the performance of the main tasks. 

Our work explores a combination of parallel and auxiliary tasks. Additionally, our feature space consists of user, product and query features which results in a combination of tabular and textual features. Given majority of our data is tabular in nature we experimented with tabular learning methods such as the tree based XGBoost\cite{chen2016xgboost}, and more recent deep learning based methods: deep crossnet V2 (DCN-V2)\cite{wang2021dcn}, feature tokenizer-transformer (FT-T) \cite{gorishniy2021revisiting} and TabNet\cite{arik2021tabnet} and compared them with the popular multi-task learning recommendation paradigm Multi-gate Mixture of Experts (MMoE)\cite{ma2018modeling}. Additionally, to handle text features we utilized TinyBERT\cite{jiao2020tinybert} due to its size and inference speed. Lastly, we focused on a hard parameter or shared bottom approach \cite{he2022metabalance,liu2022multi,yang2023adatask} in combination with expert sharing as in MMoE \cite{ma2018modeling}. A shared bottom approach refers to shared bottom layers that extract the same
information for different tasks while the task-specific top layers are trained individually. On the other hand, expert sharing refers to employing multiple expert networks to extract knowledge from the shared bottom layer.

\section{Multi-task Product Search Ranking Model Framework }

\subsection{Problem Formulation}\label{prob_form}

In the context of personalized web product search ranking, the objective is to present users with a list of products that not only match the search query but also align with the user's personal preferences and historical behavior. This problem can be formulated as a multi-task learning (MTL) problem, where multiple related tasks are optimized simultaneously to enhance overall relevance and personalization.

The input to the model consists of three main components: the query (\(Q\)), the product/item (\(I\)), and the customer (\(C\)). The query is represented as a text. The product is represented by both textual features, such as product descriptions and titles, and tabular features, including categorical and numerical attributes like price, category, and brand. The customer is represented by demographic information, browsing history, and past interactions, all encoded as categorical and numerical features. In addition, we add query product interaction level numerical features to capture the query product relevance understanding. 

For each observed impression of a product given a query, we have three binary labels: Click (\(Y_1\)), Add to Cart (ATC) (\(Y_2\)), and Purchase (TRX) (\(Y_3\)). Each label indicates whether the corresponding action was taken by the user. The goal is to predict the likelihood of each of these actions for each product in the search results, making this a multi-task classification problem. The MTL framework simultaneously optimizes for all three tasks. These tasks share a common representation derived from the input features, which is then specialized through task-specific layers. The loss function for this MTL setup is a weighted sum of the individual losses for each task:

\begin{equation}
\label{eq:mtl_loss}
\mathcal{L} = \alpha_1 \mathcal{L}_{\text{Click}}(Y_1, \hat{Y}_1) + \alpha_2 \mathcal{L}_{\text{ATC}}(Y_2, \hat{Y}_2) + \alpha_3 \mathcal{L}_{\text{TRX}}(Y_3, \hat{Y}_3)
\end{equation}

Here, \(\mathcal{L}_{\text{Click}}\), \(\mathcal{L}_{\text{ATC}}\), and \(\mathcal{L}_{\text{Trx}}\) represent the binary cross-entropy losses for predicting clicks, add-to-cart actions, and purchases, respectively. The weights \(\alpha_1\), \(\alpha_2\), and \(\alpha_3\) control the trade-off between the tasks.

By optimizing this multi-task setup, the model leverages shared information across tasks, improving overall performance and personalization in product ranking.

\subsection{Domain Specific Query Product semantic score}\label{sem_emb}
To enhance the personalized product search ranking, we incorporate a domain-specific semantic relevance score between the query and the product. This score is derived using a fine-tuned GTE\cite{li2023towards} model, a general-purpose text embedding model trained with multi-stage contrastive learning. The GTE model is adept at capturing nuanced semantic relationships between text pairs, making it well-suited for determining the relevance of a product to a user's query. The product document is represented by a combination of its textual attributes, including the product title, brand, color, and target age group. Each of these attributes is embedded using the GTE model, which transforms them into a dense vector representation. Similarly, the user's search query is embedded into a dense vector using the same model. The semantic relevance score is then computed as the cosine similarity between the query vector and the product document vector, effectively capturing the semantic alignment between the user's intent and the product's description. By leveraging the semantic relevance score obtained from the fine-tuned GTE model, we aim to improve the ranking quality by more accurately reflecting the user's intent. This approach allows the model to prioritize products that are semantically aligned with the query, thereby enhancing both the relevance and personalization of the search results.

\subsection{Model Architecture}
\begin{figure*}[ht]
  \centering
  \begin{tabular}{@{}c@{\hspace{3mm}}c@{}@{\hspace{3mm}}c}
    \includegraphics[width=0.45\textwidth]{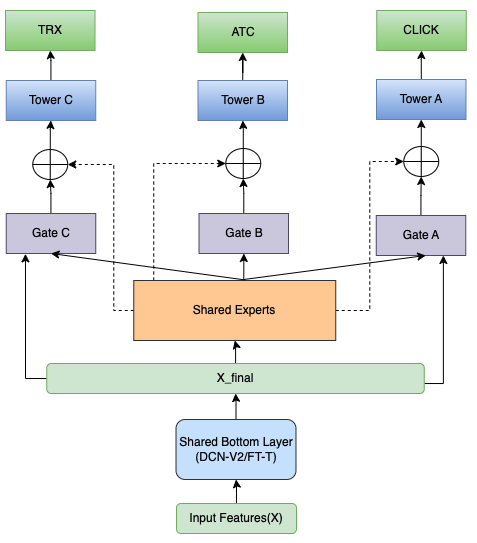} &
    
    \includegraphics[width=0.55\textwidth]{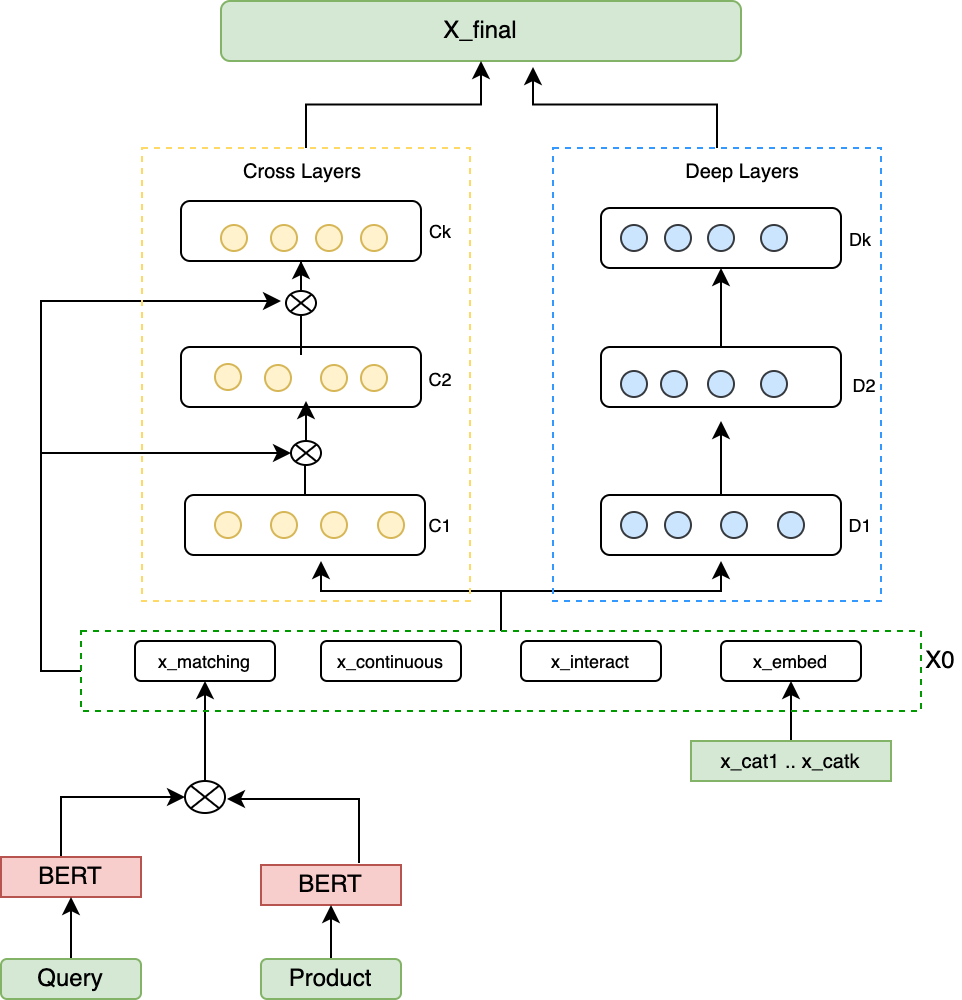} &

  \end{tabular}
  \caption{ (a) MMOE based Multi-task architecture for 3 tasks (b) DCN-V2 as shared bottom layer}
  \label{fig:mmoe_dcn_ftt}
\end{figure*}

To integrate multiple tasks into same learning objective, we adopt Multi-gate Mixture-of-Experts (MMOE)\cite{ma2018modeling} for optimizing the objective shown in  (\ref{eq:mtl_loss}). The structure of a typical MMOE architecture can be divided into following  parts: gates, experts,
towers and a shared layer at the bottom. Our main contribution is focused on shared bottom layer. We experiment with two main architectures for the shared layer: DCN-V2\cite{wang2021dcn} and FT-T\cite{gorishniy2021revisiting}. The main model architecture is shown in Fig. \ref{fig:mmoe_dcn_ftt}(a).


\subsubsection{Model Input Layer}
As described in Section \ref{prob_form}, our feature set consists of $x_{\text{categorical}}$ and $x_{\text{continuous}}$, which are independent features primarily derived from customer, query, and product data. The categorical features capture discrete properties, such as product categories and user demographics, while the continuous features represent numerical information, such as prices, ratings, and other scalar attributes.

In addition to these, we incorporate interaction features, $x_{\text{interaction}}$, which are dependent on the relationship between query and product data. These interaction features include aggregated statistics like query-item click rates over time, handcrafted features that quantify the similarity between query text and product titles, and the semantic score, which is described in more detail in Section \ref{sem_emb}. These features help model the relevance of a product to a specific query by capturing query-product interactions.

\paragraph{Categorical Embedding}\label{cat_emb} For the $i$-th categorical feature in $x_{\text{categorical}}$, we project it from a high-dimensional sparse space to a lower-dimensional dense space using the following embedding:
\begin{equation}
    \mathbf{x}^{\text{embed}}_{i} = \mathbf{W}^{\text{embed}}_{i} \mathbf{e}_i, \quad  \in \mathbb{R}^{d}
\end{equation}
where $\mathbf{e}_i \in \{0, 1\}^{v_i}$ is a one-hot encoded vector representing the categorical feature, and $\mathbf{W}^{\text{embed}}_{i} \in \mathbb{R}^{e_i \times v_i}$ is a learned projection matrix. Here, $\mathbf{x}^{\text{embed}}_{i} \in \mathbb{R}^{e_i}$ denotes the dense embedded vector, while $v_i$ and $e_i$ represent the vocabulary size and embedding size d, respectively.

\subsubsection{TinyBERT Matching Layer}
To better model the relevance between the query and the product, we incorporate a pre-trained TinyBERT model \cite{jiao2020tinybert} and fine-tune a few of its layers to extract meaningful embeddings for both the query and the product text. Inspired by the approach in \cite{qin2020matching}, we generate a feature matching representation, $\mathbf{x}_{\text{matching}}$, by performing element-wise multiplication between the query and product embeddings. This method captures the interaction between the query and product representations more effectively than a simple dot product. The process for generating the matching feature representation is as follows:
\begin{equation}
\label{eq:bert}
\mathbf{\mathcal{L}}_{\text{text}} = \text{MaxPooling}(\text{TinyBERT}(\text{text}))
\end{equation}
\begin{equation}
\label{eq:matchingop}
\mathbf{x}_{\text{matching}} = \mathbf{\mathcal{L}}_{\text{Query}} \odot \mathbf{\mathcal{L}}_{\text{Product}}
\end{equation}

We demonstrate that this element-wise multiplication provides a richer representation of the query-product relevance compared to traditional dot product methods, leading to significant gains in relevance modeling.

\subsubsection{Shared Bottom Layer with DCN-V2 architecture}
This shared bottom architecture is shown in Fig.\ref{fig:mmoe_dcn_ftt}(b). To build the input for the model using DCN-V2 \cite{wang2021dcn} as the Shared Bottom Layer, we concatenate all continuous features with the embedded representations of the categorical features, creating a combined feature vector represented as:
\begin{equation}
    \mathbf{x}_0 = [\mathbf{x}_{\text{continuous}}, \mathbf{x}_{\text{matching}}, \mathbf{x}_{\text{interaction}}, \mathbf{x}_{1}^{\text{embed}}, \ldots, \mathbf{x}_{n}^{\text{embed}}],
\end{equation}
where $\mathbf{x}_{\text{continuous}}$ represents the continuous features, $\mathbf{x}_{\text{matching}}$ and $\mathbf{x}_{\text{interaction}}$ are additional feature representations, and $\mathbf{x}_{i}^{\text{embed}}$ denotes the embedding of the $i$-th categorical feature. 

\textit{\textbf{Cross Layer}} in DCN-V2 is designed to model explicit feature interactions. The transformation at each cross layer is defined as:
\begin{equation}
\mathbf{f}^{l}_{\text{cross}}({h}_{n+1}) = \mathbf{h}_0 \odot (\mathbf{W}_n \mathbf{h}_n + \mathbf{b}_n) + \mathbf{h}_n, \quad \text{for } n = 0, 1, \ldots, l-1.
\label{eq:cross_layer_func}
\end{equation}
where $\mathbf{h}_n$ is the output of the $n$-th cross layer, $\mathbf{h}_0$ is the original input vector, $\mathbf{W}_n$ and $\mathbf{b}_n$ are the learnable weight matrix and bias vector for the $n$-th layer, respectively, and $\odot$ denotes element-wise multiplication.

\textit{\textbf{Deep Layer}} is designed to model complex, non-linear feature interactions. The transformation at each deep layer is defined as:
\begin{equation}
\mathbf{f}^{l}_{\text{deep}}({h}_{n+1}) = \text{ReLU}(\mathbf{W}_n \mathbf{h}_n + \mathbf{b}_n), \quad \text{for } n = 0, 1, \ldots, l-1.
\label{eq:deep_layer_func}
\end{equation}

where $\mathbf{h}_n$ is the output of the $n$-th deep layer, and $\mathbf{W}_n$ and $\mathbf{b}_n$ are the learnable weight matrix and bias vector for the $n$-th layer, respectively.

\textit{\textbf{Final Output Layer of DCN-V2}}: Following the parallel structure of DCN-V2\cite{wang2021dcn}, this combined input vector $\mathbf{x}_0$ is simultaneously fed into both the cross network and the deep network. The final output is obtained by concatenating the outputs of these two networks, as given by:
\begin{equation}
\mathbf{x}_{\text{final}} = [f^{L_{\text{cross}}}_{\text{cross}}(\mathbf{x}_0), f^{L_{\text{deep}}}_{\text{deep}}(\mathbf{x}_0)],
\end{equation}
where $f^{L_{\text{cross}}}_{\text{cross}}$ and $f^{L_{\text{deep}}}_{\text{deep}}$ represent the transformations applied by the cross network and deep network, respectively, and $L_{\text{cross}}$ and $L_{\text{deep}}$ represent the total number of cross and deep layers.

\subsubsection{Shared Bottom Layer with FT-Transformer architecture}
\begin{figure}[ht]
    \centering
    \includegraphics[width=0.8\columnwidth]{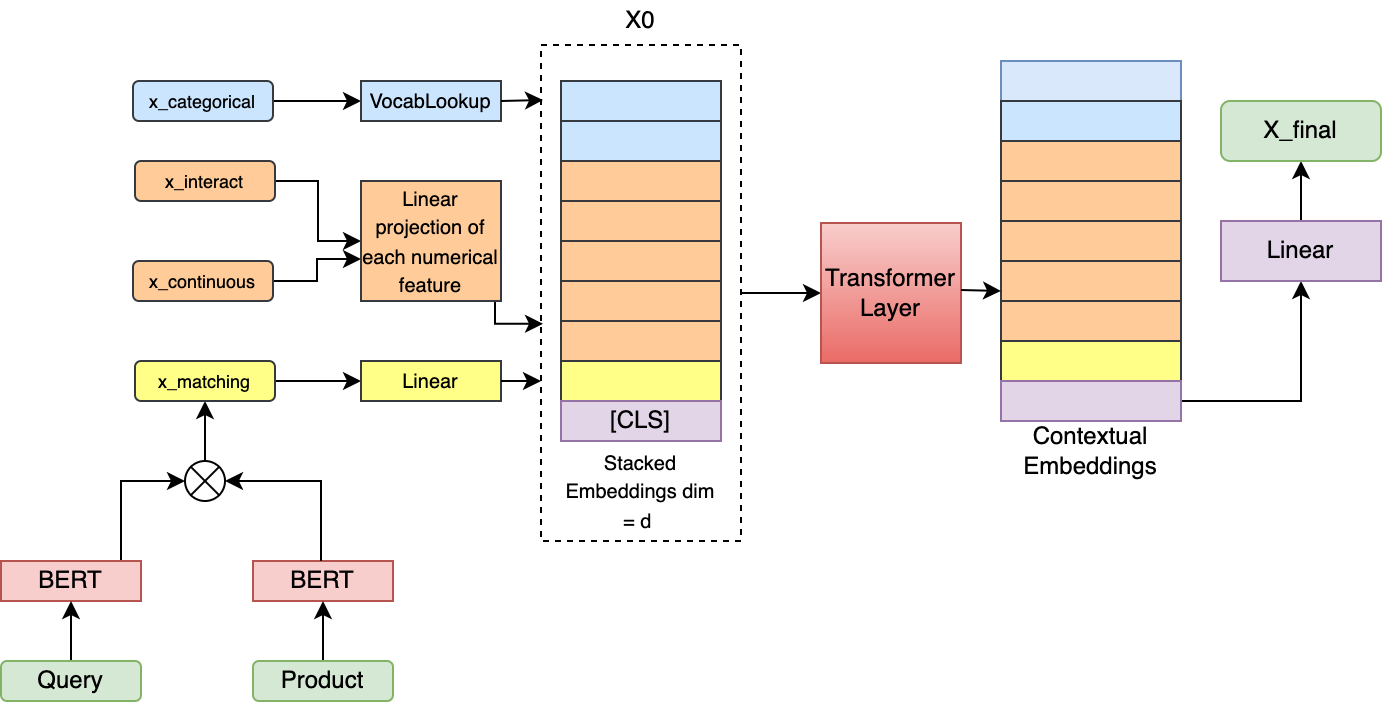} 
    \caption{ FT-T as shared bottom layer}
    \label{fig:ftt}
\end{figure}

The core concept of this bottom layer is to leverage transformer architecture for modeling tabular data. The shared bottom architecture for this approach is depicted in Fig.\ref{fig:ftt}.

\textit{\textbf{Feature Tokenizer}}: The categorical features are embedded using the same transformation as described in \ref{cat_emb}. The numerical features are transformed into dense vectors as follows:  For the $i$-th numerical feature in $x_{\text{continuous}}$,$x_{\text{interaction}}$ we project it to an embedding space using element-wise multiplication as follows:
\begin{equation}
\mathbf{x}^{\text{num}}_{i} = \mathbf{W}^{\text{num}}_{i} \cdot \mathbf{x}_i + \mathbf{b}_i, \quad  \in \mathbb{R}^{d}
\end{equation}
where $\mathbf{x}_i $ is the numerical feature, and $\mathbf{W}^{\text{num}}_{i} \in \mathbb{R}^{d}$ denoting a separate embedding layer per $i^th$ numeric feature.
\begin{equation}
\mathbf{x}^{\text{embed}}_{\text{matching}} = \mathbf{W}_m \cdot \mathbf{x}_{\text{matching}} + \mathbf{b}_m, \in \mathbb{R}^{d}
\end{equation}

For the cross product output $x_{\text{matching}}$  from Eq-\ref{eq:matchingop}, we reduce it to the same dimension $d$ as the rest of the features' embedding dimensions.

\textit{\textbf{Final output using Transformer Layer}}: The final input to the transformer layer is constructed by stacking the [CLS] token, numerical features, and categorical feature embeddings. This can be expressed as:
\begin{equation}
x_0 = \text{STACK}\left( \text{[CLS]}, x^{(\text{num})}_{1}, \ldots, x^{(\text{num})}_{k_{\text{num}}}, x^{(\text{embed})}_{1}, \ldots,  x^{(\text{embed})}_{k_{\text{embed}}}, x^{\text{embed}}_{\text{matching}} \right)
\end{equation}

Here, the [CLS] token, adapted from the NLP domain, is concatenated with the embeddings of categorical and numerical features. The combined input, \(x_0\), is then processed through multiple Transformer blocks to create a contextualized representation.

After the Transformer layers, the contextualized embedding of the [CLS] token is used as the input to a linear layer to produce the final output of the shared bottom layer, with FT-T as the architecture block. This process is mathematically represented as:
\begin{equation}
\mathbf{x}_{\text{final}} = \text{Linear}(\text{ReLU}(\text{LayerNorm}(T_{\text{[CLS]}}^{L})))
\end{equation}
In this expression, \(T_{\text{[CLS]}}^{L}\) represents the output obtained after passing the input \(x_0\) through \(L\) Transformer layers. It denotes the final representation of the [CLS] token from the \(L\)-th layer.
\vspace{-5mm}
\subsubsection{Expert and Gating Networks}
MMOE relies on expert networks to decipher the influence of various input features and employs gating functions to combine these expert outputs as proposed in \cite{ma2018modeling}.
\paragraph{Gating Networks} For each task, we create a corresponding $Softmax()$-based gating network:
\begin{equation}
\label{output:gate}
G^t(x) = \text{Softmax}(W^T x_{\text{final}})
\end{equation}
where $W^T \in \mathbb{R}^{n \times d}$ is the parameter matrix for task $t$, $n$ is the number of shared experts, and $d$ is the input dimension of $x_\text{final}$, which is the output of the shared bottom layer.

\paragraph{Expert Networks} The experts are shared across all tasks and are added on top of the shared bottom layer. We set the number of experts in our model greater than the number of tasks. Here, $E_i$ is the $i$-th expert network, and $t$ represents the task.
\begin{equation}
\label{output:gate_experts}
H^t(x) = \sum_{i=1}^{n} G^t(x) E_i(x_{\text{final}})
\end{equation}

\subsubsection{Tower Networks}
The tower architecture of our model consists of three towers, each associated with a specific task. Each tower has its own set of input gates, which are responsible for selecting the appropriate experts to handle the input data as described above. Each tower network is a simple multi-layer feed-forward network, and it can be extended to any advanced structure. The final output from the tower is obtained as follows:
\begin{equation}
\label{output:tower}
Y^t = \text{MLP}^t(H^t(x_{\text{final}}))
\end{equation}

\subsection{Data Sampling}\label{data_samp}
To ensure that our training data accurately reflects the diverse purchasing behaviors of customers across various product categories while addressing the skew towards negative samples, we employ a sophisticated sampling strategy. This approach focuses on product popularity within categories and balances positive and negative interactions. Impressions where customers interacted, like clicks, ATC, or TRX, are treated as positive, and the rest are negative.

Let the entire training dataset $D$ be composed of $c$ different subsets $\{D_1, \ldots, D_c\}$, where each subset $D_i$ corresponds to a specific product category. Our sampling strategy proceeds as follows:
\begin{enumerate}
    \item For each category $D_i$, we divide the products into $x$ bins based on the number of impressions (i.e., how often the product was viewed by customers). Let $B_{i,k}$ represent the $k$-th bin of category $D_i$, where $k \in \{1, \ldots, x\}$.
    \item The size of each bin $B_{i,k}$ is denoted as $|B_{i,k}|$.
    \item We calculate the sampling size for each bin $n_{i,k}$ as: $n_{i,k} = \beta \cdot |B_{i,k}|$, where $\beta \in (0,1]$ is a hyperparameter that controls the overall sampling rate.
    \item To address the dataset skew towards negative samples, we further divide each bin $B_{i,k}$ into positive and negative samples. Let $B_{i,k}^+$ and $B_{i,k}^-$ represent the positive and negative samples in bin $k$ of category $i$, respectively.
    We calculate the sampling sizes for positive and negative samples as: $ n_{i,k}^+ = \alpha \cdot n_{i,k} \quad \text{and} \quad n_{i,k}^- = (1-\alpha) \cdot n_{i,k}$, where $\alpha \in (0,1)$ is a hyperparameter that controls the balance between positive and negative samples.
    \item We sample $n_{i,k}^+$ positive samples from $B_{i,k}^+$ and $n_{i,k}^-$ negative samples from $B_{i,k}^-$ for each bin.
\end{enumerate}

This strategy helps the model learn patterns relevant to each category while accounting for variations in product popularity and the imbalance between positive and negative samples. By sampling from different impression-based bins within each category and controlling the ratio of positive to negative samples, we capture the full spectrum of product popularity within each category. 

This approach ensures that both frequently and infrequently viewed products are represented in our training data and the model learns from a diverse set of customer-product interactions. The skew towards negative samples is addressed, allowing for a more balanced representation of positive and negative interactions.

By carefully tuning the hyperparameters $\beta$ and $\alpha$, we can control the overall sampling rate and the trade-off between maintaining the original data distribution and creating a more balanced dataset for training. This sampling strategy ultimately allows the model to learn a more robust and generalizable representation of customer preferences across various product categories and popularity levels.

\subsection{Relevance Label Generation}\label{rel_label}

To enhance our model's performance, we introduce a fourth optimization task focused on a relevance label that quantifies the pertinence of a document to a given query. While human judgments are ideal for creating such labels, they are often prohibitively expensive and challenging to scale. As an alternative, we leverage a combination of click-feedback data and semantic similarity, offering a more scalable and comprehensive approach. Our methodology for generating relevance labels is inspired by the work of \cite{agrawal2009generating,yao2021learning}, which utilizes a multi-faceted voting mechanism. This mechanism synthesizes four key features:
\begin{enumerate}
    \item Click-through rate (CTR): We calculate a weighted CTR that accounts for both the frequency of clicks and their quality.
    \item Click position: We incorporate the position of clicked products in the search results, applying a logarithmic decay function to assign higher weight to clicks on higher-ranked items.
    \item Transaction weight: We factor in the number of transactions (purchases) resulting from clicks, adding a weight to clicks that led to transactions.
    \item Semantic similarity: We compute a semantic similarity score between the query and the document using a pre-trained semantic model.
\end{enumerate}
The process of generating relevance labels proceeds as follows:
\begin{enumerate}
    \item We aggregate impression, click, and transaction data for each query-product pair across all positions.
    \item We compute a position weight using a logarithmic function: $\text{position\_weight} = \log(\text{position} + 1)^{1.5}$. This gives more importance to clicks on higher-ranked items while still valuing lower-ranked clicks.
    \item We calculate a transaction weight as: $\text{transaction\_weight} = 1 + \frac{\text{transactions}}{\text{clicks}}$. This amplifies the importance of clicks that led to purchases.
    \item We then compute weighted clicks as: $\text{weighted\_clicks} = \text{clicks} \times \text{position\_weight} \times \text{transaction\_weight}$.
    \item The weighted CTR is calculated as: $\text{weighted\_CTR} = \frac{\text{weighted\_clicks}}{\text{impressions}}$.
    \item We compute a semantic similarity score $\text{sem\_score}$ between the query and the document using our pre-trained semantic model, as described in Section \ref{sem_emb}.
    \item We combine the weighted CTR and semantic score into a single relevance score:
       $\text{relevance\_score} = \alpha \cdot \text{normalized(weighted\_CTR)} + (1-\alpha) \cdot \text{sem\_score}$
       where $\alpha$ is a hyperparameter that controls the balance between click-based and semantic-based relevance.
\end{enumerate}

This approach allows us to generate nuanced relevance labels that account for multiple aspects of user behavior, including click frequency, click position, and purchase decisions, as well as the semantic relationship between queries and documents. By using query-specific thresholds, we ensure that the relevance labels are contextually appropriate for each query.

The incorporation of the semantic similarity score adds an important dimension to our relevance assessment. While click-based metrics capture user behavior and preferences, they may be biased by factors such as item position or visual appeal. The semantic score provides a content-based relevance measure that can help mitigate these biases and capture deeper relationships between queries and documents.

For model optimization, we treat this as a multi-class classification problem and employ categorical cross-entropy loss. The impact of including this relevance label task in our model training is further examined in our ablation study, where we analyze its contribution to overall model performance and generalization.

\subsection{Other Models}\label{comp_models}
In this section, we outline other models used for comparison in optimizing personalized product search ranking. Some of these models are specifically designed for tabular data, while others can handle both tabular and non-tabular data. Our proposed model architecture is capable of incorporating non-tabular data effectively.\\
\textbf{XGBoost} \cite{chen2016xgboost}: XGBoost is a gradient boosting framework that uses tree-based learning algorithms. It has been widely applied in learning-to-rank systems but is limited to handling only tabular data.\\
\textbf{TabNet} \cite{arik2021tabnet}: TabNet is a model that employs a recurrent architecture with sequential attention mechanisms at each decision step to select features. It is specifically designed for tabular data.\\
\textbf{Feature Tokenizer Transformer (FT-T)} \cite{gorishniy2021revisiting}: This model uses a transformer layer to contextualize categorical and numerical features through a feature tokenizer encoder. While the original architecture is flexible enough to accept non-tabular data, it is not ideal due to its high computational resource requirements for training.\\
\textbf{DCN-V2} \cite{wang2021dcn}: The DCN-V2 model introduces a cross layer that captures feature interactions, allowing it to work with both tabular and non-tabular data. Its main contribution is the efficient modeling of feature interactions.\\
\textbf{MMoE (Multi-gate Mixture-of-Experts)} \cite{ma2018modeling}: MMoE integrates multiple experts through a gating mechanism to control the contribution of each expert. It has been effectively used for modeling ranking tasks and supports both tabular and non-tabular data.

\section{Experiments}

\subsection{Dataset}\label{AA}
\vspace{-2mm}
We performed experiments on proprietary data containing users, products, search queries, and their corresponding click, add-to-cart, and transaction interactions. The training dataset was created using the sampling technique described in Section \ref{data_samp}. The data was split into training, validation, and test sets with proportions of 80\%, 10\%, and 10\%, respectively. Table \ref{tab:dataset_stats} presents the statistics for the sampled dataset.
\vspace{-5mm}
\begin{table}[htbp]
\caption{Dataset statistics}
\centering
\small
\begin{tabular}{|c|c|c|c|}
\hline
\textbf{Queries} & \textbf{Products} & \textbf{Users} & \textbf{Impressions} \\
\hline
7,181,959 & 3,619,723 & 18,054,693 & 59,153,886 \\
\hline
\end{tabular}
\label{tab:dataset_stats}
\end{table}
\vspace{-10mm}

\subsection{Training Details}\label{training_dets}
\vspace{-2mm}
The models were trained using the Adam optimizer with a learning rate of 0.001 and a batch size of 16384. The training process spanned 10 epochs, and early stopping was implemented based on validation loss to prevent overfitting. The loss weights in the multi-task learning framework were tuned using grid search, exploring values within the range of 0 to 1 with a step size of 0.1. Hyperparameter tuning for other key parameters, such as dropout rates and layer sizes, was performed using random search. The training was conducted on a high-performance computing cluster equipped with NVIDIA Tesla A100 GPUs. 
\vspace{-3mm}

\subsection{Evaluation Metrics}
\vspace{-2mm}
To evaluate the effectiveness of the various methods on the pointwise tasks of click prediction, add-to-cart (ATC) prediction, and transaction prediction, we use the AUC-ROC metric. This metric provides a robust measure of model performance by evaluating the ability to distinguish between different classes, which is crucial for our ranking use-case. For assessing the quality of ranking, we use the MRR@K. The MRR@K metric captures the sensitivity of users to the ranked positions of products in the search results, reflecting the relevance of the top-ranked items in response to a search query.

In addition to standard metrics, we introduce a new metric, \textit{Personalization Degree} (PD), to quantify the impact of personalization by measuring the variance in ranking results. 
Here are the steps to compute PD:
\begin{itemize}
    \item Randomly sample a subset \(S\) from the test dataset, denoted as \(Test_p\).
    \item Generate a modified dataset, \(Test_m\), by setting all user-specific numerical features to 0 and categorical features to default values for each sample in \(Test_p\).
    \item Compute the top-K ranked items for each query in both \(Test_p\) and \(Test_m\).
    \item Define PD as the average intersection of the top-K results between \(Test_p\) and \(Test_m\).
\end{itemize}
The metric \(PD@K\) is defined as:
\small
\begin{equation}
PD@K = \frac{1}{|S|} \sum_{i=1}^{|S|} \frac{|TopK(Test_p^i) \cap TopK(Test_m^i)|}{K}
\label{eq:personalization_degree}
\end{equation}
\normalsize
where a lower intersection indicates a higher degree of personalization, reflecting greater influence of user-specific features on the rankings.

\section{Experimental Results}

\subsection{Baseline comparison}
We compare our proposed model against the baseline variations outlined in Section \ref{comp_models}, with the experimental results summarized in Table \ref{tab:model_performance}. To compute the final ranked positions of products for MRR, we optimize the loss weights for the three tasks by performing hyperparameter tuning as described in Section \ref{training_dets}. The number of trainable and total parameters in the model provide insight into the model's complexity and potential for overfitting. The table compares the performance of different models, including single-task and multi-task models, with and without the incorporation of TinyBERT embeddings. Multi-task learning generally outperforms single-task learning, MMoE Models which optimize multiple tasks simultaneously, tend to achieve better performance across various metrics compared to single-task models like XGBoost, DCN-V2, and FT-T. Models that leverage TinyBERT for semantic embeddings generally exhibit superior performance compared to their counterparts without TinyBERT. The performance varies depending on whether DCN-V2 or FT-T is used as the shared bottom layer in the MMoE framework.  
\vspace{-5mm}
\begin{table*}[!htbp] 
\caption{Performance metrics and model characteristics.}
\begin{adjustbox}{width=1.0\textwidth}
{
\small
\begin{tabular}{|l|ccc|ccc|ccc|c|cc|}
\hline
\textbf{Model Name} & \multicolumn{3}{c|}{\textbf{AUC-ROC}} & \multicolumn{3}{c|}{\textbf{MRR@1}} & \multicolumn{3}{c|}{\textbf{Task Weights}} & \textbf{PD@10} & \multicolumn{2}{c|}{\textbf{\# Parameters}} \\
\cline{2-13}
 & \textbf{Click} & \textbf{ATC} & \textbf{TRX} & \textbf{Click} & \textbf{ATC} & \textbf{TRX} & \textbf{Click} & \textbf{ATC} & \textbf{TRX} &  & \textbf{Trainable} & \textbf{Total}  \\
\hline
XGBoost (Single ATC Task) & - & 0.869 & - & 0.302 & 0.364 & 0.377 & - & - & - & - & - & - \\
DCN-V2 (Single ATC Task) & - & 0.867 & - & 0.302 & 0.368 & 0.385 & - & - & - & - & 850K & 850K\\
FT-T (Single ATC Task) & - & 0.863 & - & 0.347 & 0.376 & 0.390 & - & - & - & - & 927K & 927K\\
MMoE & 0.726 & 0.784 & 0.776 & 0.277 & 0.325 & 0.342 & 0.2 & 0.7 & 0.1 & - & 100K & 100K\\
TabNet & 0.834 & 0.872 & 0.859 & \textbf{0.354} & 0.372 & 0.387 & 0.1 & 0.2 & 0.7 & - & 580K & 580K\\
FT-T & 0.764 & 0.858 & 0.844 & 0.353 & 0.377 & 0.393 & 0.1 & 0.9 & 0.0 & \textbf{0.87} & 927K & 927K\\
MMoE with shared bottom DCN-V2 & 0.830 & 0.868 & 0.855 & 0.329 & 0.375 & 0.391 & 0.4 & 0.1 & 0.5 & 0.92 & 998K & 998K\\
MMoE with shared bottom FT-T & 0.819 & 0.855 & 0.846 & 0.326 & 0.375 & 0.389 & 0.3 & 0.4 & 0.3 & - & 2.9M & 2.9M\\
MMoE with shared bottom DCN-V2 + TinyBERT Cross Layer & 0.824 & 0.863 & 0.850 & 0.324 & 0.373 & 0.391 & 0.4 & 0.3 & 0.3 & 0.88 & 9.6M & 69.5M\\
MMoE with shared bottom FT-T + TinyBERT Cross Layer & \textbf{0.838} & 0.862 & \textbf{0.864} & 0.352 & \textbf{0.378} & \textbf{0.395} & 0.5 & 0.3 & 0.2 & \textbf{0.87} & 10.1M & 70M\\

\hline
\end{tabular}}
\end{adjustbox}
\label{tab:model_performance}
\end{table*}
\vspace{-5mm}

Overall, Table-\ref{tab:model_performance} provides a comprehensive overview of the performance and characteristics of various models, offering valuable insights into the factors that contribute to effective personalized product search ranking. The results emphasize the benefits of multi-task learning, the incorporation of semantic embeddings, and the importance of careful task weight tuning and model selection in achieving optimal performance. Integrating MMoE, FT-T, and TinyBERT-based cross layer techniques can significantly improve overall MRR performance in personalized product search ranking. MMoE with shared bottom FT-T with TinyBERT Cross Layer model achieves the highest MRR scores for Add-to-Cart (0.378) and Transaction (0.395) tasks, while maintaining competitive performance for the Click task (0.352). This architecture effectively combines the strengths of MMoE's task-specific expert learning, FT-T's ability to handle both categorical and numerical features, and TinyBERT's advanced semantic understanding of textual data. The synergy of these components allows the model to capture complex interactions between user intent, product attributes, and contextual factors, resulting in more relevant and personalized search rankings. While this approach does increase the total number of model parameters, the substantial improvements in ranking quality suggest that the added complexity is justified by the enhanced user experience and potential for increased conversions in e-commerce applications. 
We have also shown PD@10 scores for few models, it ranges from 0.87 to 0.92, signifying that personalization alters approximately one in top ten product recommendations because of user-specific features. PD@10 helps evaluate how effectively a model tailors search results to individual users. A lower PD@10 signifies that the model is successfully leveraging user data to provide a more personalized search experience. 


\vspace{-2mm}
\subsection{Ablation Study}
We limit the ablation study to variations of the DCN-V2 model in order to optimize computational bandwidth, as running it on all model variations would be resource-intensive.

\vspace{-5mm}
\subsubsection{Impact of TinyBERT-Fine tuning layer} To assess the impact of the TinyBERT-based layer, we conduct an ablation study by removing the semantic score feature described in Section \ref{sem_emb}. The results, shown in Table \ref{tab:model_performance2}, demonstrate that the model with the TinyBERT\cite{jiao2020tinybert} fine-tuning layers retains its overall performance even showing minor improvements in click MRR compared to Table \ref{tab:model_performance}. In contrast, models like TabNet and MMoE + DCN-V2 experienced significant performance declines, with reductions in all MRR metrics, compared to their results in Table \ref{tab:model_performance}.

This analysis highlights a practical advantage: removing the costly semantic score feature allows the model to maintain competitive results while avoiding the overhead of maintaining two large models. For example, in the case of TinyBERT\cite{jiao2020tinybert}, there is no performance drop, making it a more efficient solution in terms of both computation and resource allocation, without sacrificing substantial ranking quality.
\vspace{-5mm}
\begin{table}[H]  
\caption{Comparison of model performance without the semantic score feature, showing improvements relative to Table \ref{tab:model_performance}.}
\centering
\scriptsize
\setlength{\tabcolsep}{4pt} 
\renewcommand{\arraystretch}{1.3} 
\begin{tabularx}{\columnwidth}{|X|ccc|ccc|} 
\hline
\textbf{Model Name} & \multicolumn{3}{c|}{\textbf{MRR@1}} & \multicolumn{3}{c|}{\textbf{Task Weights}} \\
\cline{2-7}
 & \textbf{Click} & \textbf{ATC} & \textbf{TRX} & \textbf{Click} & \textbf{ATC} & \textbf{TRX} \\
\hline
TabNet & \makecell{0.347 (-1.97\%)} & \makecell{0.368 (-1.07\%)} & \makecell{0.380 (-1.8\%)} & 0.1 & 0.9 & 0.0  \\
MMoE + DCN-V2 & \makecell{0.322 (-2.12\%)} & \makecell{0.374 (-0.26\%)} & \makecell{0.389 (-0.51\%)} & 0.3 & 0.3 & 0.4  \\
MMoE + DCN-V2 + TinyBERT Cross & \makecell{0.328  (+1.23\%)} & \makecell{0.373 (0.00\%)} & \makecell{0.391  (0.00\%)} & 0.5 & 0.1 & 0.4  \\
\hline
\end{tabularx}
\label{tab:model_performance2}
\end{table}
\vspace{-10mm}

\subsubsection{Cross vs Dot Product for the TinyBERT matching Layer}
Based on the results in Table \ref{tab:bert_cross_vs_dot}, the use of element-wise multiplication (Cross) in our approach provides a more effective encoding of query-product relevance compared to the standard dot product method. This is reflected in the improved performance across all tasks (Click, ATC, and TRX), where the model utilizing the cross product consistently outperforms the dot product approach. Specifically, we observe slight but meaningful gains in MRR@1 for all tasks, with the cross product model achieving an increase of approximately ($+0.91\%$) for click prediction, ($+0.53\%$) for add-to-cart prediction, and ($+0.77\%$) for transaction prediction compared to the dot product. These percentage improvements, though modest, indicate that the cross product provides a more effective and accurate relevance modeling framework.
\vspace{-5mm}
\begin{table}[H]
\caption{Cross vs Dot Product for the TinyBERT matching Layer}
\centering
\scriptsize
\begin{tabular}{|p{4.5cm}|ccc|ccc|} 
\hline
\textbf{Model Name} & \multicolumn{3}{c|}{\textbf{MRR@1}} & \multicolumn{3}{c|}{\textbf{Task Weights}} \\
\cline{2-7}
 & \textbf{Click} & \textbf{ATC} & \textbf{TRX} & \textbf{Click} & \textbf{ATC} & \textbf{TRX} \\
\hline
\makecell[l]{MMoE + DCN-V2 + TinyBERT Dot} & 0.320 & 0.371 & 0.388 & 0.4 & 0.2 & 0.4   \\
\makecell[l]{MMoE + DCN-V2 + TinyBERT Cross} & \textbf{0.324} & \textbf{0.373} & \textbf{0.391} & 0.4 & 0.3 & 0.3  \\
\hline
\end{tabular}
\label{tab:bert_cross_vs_dot}
\end{table}

\vspace{-10mm}
\subsubsection{Performance with addition of Relevance Label Task}
To investigate the effect of task expansion, we introduced a fourth task - relevance label prediction - while maintaining the core architecture (Table~\ref{tab:model_performance}). This augmented model demonstrated improved performance across all existing tasks, with MRR@1 scores of 0.324 ($+1.22\%$) for click prediction, 0.373 ($+0.57\%$) for add-to-cart prediction, and 0.391 ($+0.86\%$) for transaction prediction. Notably, the click prediction task exhibited the most substantial improvement, suggesting that the inclusion of the relevance label task particularly enhances the model's ability to predict user clicks. These results indicate that the addition of a related task can lead to positive transfer learning, potentially due to shared underlying patterns between relevance and user engagement metrics.
\vspace{-5mm}
\begin{table}[H]
\caption{Performance with Relevance Label Task}
\centering
\scriptsize
\begin{tabular}{|p{4.5cm}|ccc|ccc|} 
\hline
\textbf{Model Name} & \multicolumn{3}{c|}{\textbf{MRR@1}} & \multicolumn{3}{c|}{\textbf{Task Weights}} \\
\cline{2-7}
 & \textbf{Click} & \textbf{ATC} & \textbf{TRX} & \textbf{Click} & \textbf{ATC} & \textbf{TRX} \\
\hline
MMoE + DCN-V2 + without relevance task & 0.329 & 0.375 & 0.391 & 0.4 & 0.1 & 0.5  \\
MMoE + DCN-V2 + with relevance task & \textbf{0.333} & \textbf{0.376} & \textbf{0.392} & 0.5 & 0.2 & 0.3  \\
\hline
\end{tabular}
\label{tab:perf_with_relevance_label}
\end{table}

\section{Conclusion and Future Work}
In this paper, we proposed a novel multi-task learning (MTL) framework for personalized product search ranking that effectively integrates both tabular and non-tabular data. By leveraging a pre-trained TinyBERT\cite{jiao2020tinybert} model for semantic embedding extraction and combining it with traditional feature representations, our approach enhances the model's ability to capture complex interactions between user intent, product attributes, and contextual factors. 

We conducted a comprehensive evaluation of our proposed model against several baseline models, and demonstrate that the proposed framework significantly improves performance across multiple tasks, as evidenced by superior AUC-ROC and MRR@K scores. Additionally, we introduced a novel metric, \textit{Personalization Degree}, to quantify the effectiveness of personalized search results, providing deeper insights into the impact of personalization on user experience.

Our findings indicate that incorporating non-tabular data, such as textual features through TinyBERT-based embeddings, along with carefully tuned task weights, can greatly enhance the overall ranking quality and user satisfaction in e-commerce search systems. Moreover, the relevance labeling mechanism introduced in this work offers a scalable alternative to human-annotated labels, reducing costs while maintaining robustness against noisy click logs.

Future research directions involve exploring more sophisticated embedding techniques and enhancing the relevance labeling mechanism by incorporating large language models as additional voting components for determining relevance labels. Furthermore, to improve personalization, our proposed architecture could be extended to include more dynamic, session-based user data, allowing for a more tailored user experience. Overall, this work provides a significant step forward in developing more effective and efficient personalized product search ranking models, offering a robust foundation for future advancements in the field.

\bibliographystyle{splncs04}
\bibliography{search}

\begin{thebibliography}{10}
\providecommand{\url}[1]{\texttt{#1}}
\providecommand{\urlprefix}{URL }
\providecommand{\doi}[1]{https://doi.org/#1}

\bibitem{agrawal2009generating}
Agrawal, R., Halverson, A., Kenthapadi, K., Mishra, N., Tsaparas, P.: Generating labels from clicks. In: Proceedings of the Second ACM International Conference on Web Search and Data Mining. pp. 172--181 (2009)

\bibitem{arik2021tabnet}
Arik, S.{\"O}., Pfister, T.: Tabnet: Attentive interpretable tabular learning. In: Proceedings of the AAAI conference on artificial intelligence. vol.~35, pp. 6679--6687 (2021)

\bibitem{bi2019leverage}
Bi, K., Teo, C.H., Dattatreya, Y., Mohan, V., Croft, W.B.: Leverage implicit feedback for context-aware product search. arXiv preprint arXiv:1909.02065  (2019)

\bibitem{burges2010from}
Burges, C.J.: From ranknet to lambdarank to lambdamart: An overview. Tech. Rep. MSR-TR-2010-82 (June 2010), \url{https://www.microsoft.com/en-us/research/publication/from-ranknet-to-lambdarank-to-lambdamart-an-overview/}

\bibitem{pmlr-v14-chapelle11a}
Chapelle, O., Chang, Y.: Yahoo! learning to rank challenge overview. In: Chapelle, O., Chang, Y., Liu, T.Y. (eds.) Proceedings of the Learning to Rank Challenge. Proceedings of Machine Learning Research, vol.~14, pp. 1--24. PMLR, Haifa, Israel (25 Jun 2011), \url{https://proceedings.mlr.press/v14/chapelle11a.html}

\bibitem{chen2016xgboost}
Chen, T., Guestrin, C.: Xgboost: A scalable tree boosting system. In: Proceedings of the 22nd acm sigkdd international conference on knowledge discovery and data mining. pp. 785--794 (2016)

\bibitem{frolov2017tensor}
Frolov, E., Oseledets, I.: Tensor methods and recommender systems. Wiley Interdisciplinary Reviews: Data Mining and Knowledge Discovery  \textbf{7}(3),  e1201 (2017)

\bibitem{gorishniy2021revisiting}
Gorishniy, Y., Rubachev, I., Khrulkov, V., Babenko, A.: Revisiting deep learning models for tabular data. Advances in Neural Information Processing Systems  \textbf{34},  18932--18943 (2021)

\bibitem{ijcai2017p239}
Guo, H., TANG, R., Ye, Y., Li, Z., He, X.: Deepfm: A factorization-machine based neural network for ctr prediction. In: Proceedings of the Twenty-Sixth International Joint Conference on Artificial Intelligence, {IJCAI-17}. pp. 1725--1731 (2017). \doi{10.24963/ijcai.2017/239}, \url{https://doi.org/10.24963/ijcai.2017/239}

\bibitem{he2022metabalance}
He, Y., Feng, X., Cheng, C., Ji, G., Guo, Y., Caverlee, J.: Metabalance: improving multi-task recommendations via adapting gradient magnitudes of auxiliary tasks. In: Proceedings of the ACM Web Conference 2022. pp. 2205--2215 (2022)

\bibitem{jiao2020tinybertdistillingbertnatural}
Jiao, X., Yin, Y., Shang, L., Jiang, X., Chen, X., Li, L., Wang, F., Liu, Q.: Tinybert: Distilling bert for natural language understanding (2020), \url{https://arxiv.org/abs/1909.10351}

\bibitem{jiao2020tinybert}
Jiao, X., Yin, Y., Shang, L., Jiang, X., Chen, X., Li, L., Wang, F., Liu, Q.: Tinybert: Distilling bert for natural language understanding. In: Findings of the Association for Computational Linguistics: EMNLP 2020. pp. 4163--4174 (2020)

\bibitem{ke2017lightgbm}
Ke, G., Meng, Q., Finely, T., Wang, T., Chen, W., Ma, W., Ye, Q., Liu, T.Y.: Lightgbm: A highly efficient gradient boosting decision tree. In: Advances in Neural Information Processing Systems 30 (NIP 2017) (December 2017), \url{https://www.microsoft.com/en-us/research/publication/lightgbm-a-highly-efficient-gradient-boosting-decision-tree/}

\bibitem{10.1145/2983323.2983867}
Khabsa, M., Crook, A., Awadallah, A.H., Zitouni, I., Anastasakos, T., Williams, K.: Learning to account for good abandonment in search success metrics. In: Proceedings of the 25th ACM International on Conference on Information and Knowledge Management. p. 1893–1896. CIKM '16, Association for Computing Machinery, New York, NY, USA (2016). \doi{10.1145/2983323.2983867}, \url{https://doi.org/10.1145/2983323.2983867}

\bibitem{10.1145/2556195.2556220}
Kim, Y., Hassan, A., White, R.W., Zitouni, I.: Modeling dwell time to predict click-level satisfaction. In: Proceedings of the 7th ACM International Conference on Web Search and Data Mining. p. 193–202. WSDM '14, Association for Computing Machinery, New York, NY, USA (2014). \doi{10.1145/2556195.2556220}, \url{https://doi.org/10.1145/2556195.2556220}

\bibitem{li2023towards}
Li, Z., Zhang, X., Zhang, Y., Long, D., Xie, P., Zhang, M.: Towards general text embeddings with multi-stage contrastive learning. arXiv preprint arXiv:2308.03281  (2023)

\bibitem{liu2022multi}
Liu, J., Li, X., An, B., Xia, Z., Wang, X.: Multi-faceted hierarchical multi-task learning for recommender systems. In: Proceedings of the 31st ACM International Conference on Information \& Knowledge Management. pp. 3332--3341 (2022)

\bibitem{INR-016}
Liu, T.Y.: Learning to rank for information retrieval. Foundations and Trends® in Information Retrieval  \textbf{3}(3),  225--331 (2009). \doi{10.1561/1500000016}, \url{http://dx.doi.org/10.1561/1500000016}

\bibitem{ma2018modeling}
Ma, J., Zhao, Z., Yi, X., Chen, J., Hong, L., Chi, E.H.: Modeling task relationships in multi-task learning with multi-gate mixture-of-experts. In: Proceedings of the 24th ACM SIGKDD international conference on knowledge discovery \& data mining. pp. 1930--1939 (2018)

\bibitem{naumov2019deep}
Naumov, M., Mudigere, D., Shi, H.J.M., Huang, J., Sundaraman, N., Park, J., Wang, X., Gupta, U., Wu, C.J., Azzolini, A.G., et~al.: Deep learning recommendation model for personalization and recommendation systems. arXiv preprint arXiv:1906.00091  (2019)

\bibitem{ning2015comprehensive}
Ning, X., Desrosiers, C., Karypis, G.: A comprehensive survey of neighborhood-based recommendation methods. Recommender systems handbook pp. 37--76 (2015)

\bibitem{qin2020matching}
Qin, Z., Li, Z., Bendersky, M., Metzler, D.: Matching cross network for learning to rank in personal search. In: Proceedings of The Web Conference 2020. pp. 2835--2841 (2020)

\bibitem{wang2017deep}
Wang, R., Fu, B., Fu, G., Wang, M.: Deep \& cross network for ad click predictions. In: Proceedings of the ADKDD'17, pp.~1--7 (2017)

\bibitem{wang2021dcn}
Wang, R., Shivanna, R., Cheng, D., Jain, S., Lin, D., Hong, L., Chi, E.: Dcn v2: Improved deep \& cross network and practical lessons for web-scale learning to rank systems. In: Proceedings of the web conference 2021. pp. 1785--1797 (2021)

\bibitem{wang2023multi}
Wang, Y., Lam, H.T., Wong, Y., Liu, Z., Zhao, X., Wang, Y., Chen, B., Guo, H., Tang, R.: Multi-task deep recommender systems: A survey. arXiv preprint arXiv:2302.03525  (2023)

\bibitem{Wu_2022}
Wu, X., Magnani, A., Chaidaroon, S., Puthenputhussery, A., Liao, C., Fang, Y.: A multi-task learning framework for product ranking with bert. In: Proceedings of the ACM Web Conference 2022. p. 493–501. WWW ’22, ACM (Apr 2022). \doi{10.1145/3485447.3511977}, \url{http://dx.doi.org/10.1145/3485447.3511977}

\bibitem{xiang2010context}
Xiang, B., Jiang, D., Pei, J., Sun, X., Chen, E., Li, H.: Context-aware ranking in web search. In: Proceedings of the 33rd international ACM SIGIR conference on Research and development in information retrieval. pp. 451--458 (2010)

\bibitem{yang2023adatask}
Yang, E., Pan, J., Wang, X., Yu, H., Shen, L., Chen, X., Xiao, L., Jiang, J., Guo, G.: Adatask: A task-aware adaptive learning rate approach to multi-task learning. In: Proceedings of the AAAI conference on artificial intelligence. vol.~37, pp. 10745--10753 (2023)

\bibitem{yao2021learning}
Yao, S., Tan, J., Chen, X., Yang, K., Xiao, R., Deng, H., Wan, X.: Learning a product relevance model from click-through data in e-commerce. In: Proceedings of the Web Conference 2021. pp. 2890--2899 (2021)

\bibitem{10.1145/3219819.3219823}
Zhou, G., Zhu, X., Song, C., Fan, Y., Zhu, H., Ma, X., Yan, Y., Jin, J., Li, H., Gai, K.: Deep interest network for click-through rate prediction. In: Proceedings of the 24th ACM SIGKDD International Conference on Knowledge Discovery \& Data Mining. p. 1059–1068 (2018). \doi{10.1145/3219819.3219823}

\end{thebibliography}

\end{document}